\documentclass[amssymb, amsmath, aps, prb, showkeys, showpacs]{revtex4}% Physical Review B

\usepackage{graphicx}% Include figure files
\usepackage{docs}%
\usepackage{dcolumn}% Align table columns on decimal point
\usepackage{bm}% bold math

\begin{document}

\title{Spin relaxation in mesoscopic superconducting Al wires}
\author{Yun-Sok Shin}
\altaffiliation{Present address: Electronic Devices Group, Korea
Research Institute of Standards and Science}
\author{Hu-Jong Lee}
\email{hjlee@postech.ac.kr}
\author{Hyun-Woo Lee}
\affiliation{Department of Physics, Pohang University of Science
and Technology, Pohang 790-784, Republic of Korea }

\date{\today}

\begin{abstract}
We studied the diffusion and the relaxation of the polarized
quasiparticle spins in superconductors. To that end,
quasiparticles of polarized spins were injected through an
interface of a mesoscopic superconducting Al wire in proximity
contact with an overlaid ferromagnetic Co wire in the
single-domain state. The superconductivity was observed to be
suppressed near the spin-injecting interface, as evidenced by the
occurrence of a finite voltage for a bias current below the onset
of the superconducting transition. The spin diffusion length,
estimated from finite voltages over a certain length of Al wire
near the interface, was almost temperature independent in the
temperature range sufficiently below the superconducting
transition but grew as the transition temperature was approached.
This temperature dependence suggests that the relaxation of the
spin polarization in the superconducting state is governed by the
condensation of quasiparticles to the paired state. The spin
relaxation in the superconducting state turned out to be more
effective than in the normal state.
\end{abstract}

\maketitle

Recently the spin-dependent electron transport has been the
subject of intensive studies. The key element of the phenomenon is
to inject a current of spin-polarized conduction electrons into a
mesoscopic or nano-scale non-magnetic metal or semiconductor,
control, and detect the resulting spin state. Spin-polarized
electron can be injected from a ferromagnet (F) into the system
under
study.\cite{Johnson2,Prinz1,Datta1,Nitta1,Hammar1,Zhu1,Jedema2,Jedema3}
To realize the spin-dependent electronic conductance or
``spintronics" it is essential to obtain the accurate information
on the characteristic spin-relaxation time or length of the
injected electrons in the metallic or semiconducting system in the
presence of spin-relaxing
scattering.\cite{Johnson2,Prinz1,Datta1,Nitta1,Hammar1,Zhu1,Jedema2,Jedema3}
The spin-relaxation originates from both scattering by magnetic
impurities and spin-orbit scattering of conduction electrons, but
the relaxation due to spin-orbit scattering is dominant without
magnetic impurities. A number of studies on the spin relaxation in
metals have been done using nonlocal spin
injection,\cite{Johnson2,Jedema3,Jedema4} conduction electron spin
resonance,\cite{Lubzens1,Beuneu1,Fabian1} weak
localization,\cite{Bergmann1,Gordon1} and superconducting
tunneling
spectroscopy.\cite{Meservey1,Meservey2,Grimaldi1,Monsma1} Observed
spin relaxation rate using different techniques at room
temperature, where the electron-phonon interaction predominates
the spin-orbit scattering, reveals reasonable consistency, but it
shows a wider spread at low temperatures around liquid helium
temperature. It has been pointed out that \cite{Jedema4}, as the
impurity scattering predominates the spin-orbit scattering at low
temperatures, the measured spin relaxation rates may depend on
different measurement techniques which are sensitive to different
impurity-induced spin-orbit scattering.

Recently, the spin relaxation in a superconductor (S), both
conventional\cite{Johnson1,Vier1,Nemes1,Yamashita1,Yafet1,Sillanpaa1}
and high-$T_c$ cuprate,\cite{Vas'co,Dong,Yeh,Stroud} has attracted
much research interest in relation with the recombination
mechanism of the spin-polarized quasiparticles into the singlet
Cooper-paired state. A number of studies on the spin diffusion in
conventional superconductors, however, have revealed contradicting
results. Measurements of spin accumulation effect in F/S/F-type
bipolar spin transistors\cite{Johnson1} showed an increase of the
spin-diffusion length in superconducting Nb films as
$\lambda_{sp}(T)=\lambda_{sp}(0)/(1-T/T_c)^n$ with $1/4<n<1/2$,
with increasing temperature below the superconducting transition
temperature $T_c$. But this result was in contradiction to the
increase of the spin-relaxation rate with increasing temperature
near $T_c$ from below in superconducting Nb films and
potassium-doped fulleride (K$_3$C$_{60}$) compounds measured by
the electron spin resonance technique.\cite{Vier1,Nemes1} More
recent theoretical studies by Yamashita {\it et
al.},\cite{Yamashita1} however, indicated that the estimated
spin-diffusion length in both the superconducting state
(neglecting the charge imbalance effect) and the normal-metallic
state should be the same, implying that the spin-diffusion
characteristics should be independent of temperature in the narrow
temperature range below $T_c$.

On the other hand, studies on the influence of the spin-polarized
quasiparticle injection into high-$T_c$
cuprates\cite{Vas'co,Dong,Yeh,Stroud} have mainly been focused on
the effective suppression of the superconductivity. The sensitive
dependence of the critical current on the spin injection in a
low-carrier-density cuprate hybridized with a highly polarized
colossal magnetoresistance material is expected to open a way to
develop active three-terminal superconducting devices with a high
current gain. In addition, it is expected that the spin injection
into cuprates may provide key information on the possible roles of
the spin degrees of freedom in bringing about the high-$T_c$
superconducting order. For these purposes also clear understanding
of the spin relaxation mechanism in the cuprates is an essential
element.

In this study we injected a spin-polarized current from a
ferromagnetic Co wire into a mesoscopic superconducting Al wire
which was in proximity contact with the Co wire and observed the
resulting suppression of the superconductivity in the Al wire. In
general, the superconductivity suppresses as superconducting pairs
are broken by the injection of the nonequilibrium quasiparticles
into a superconductor. In our study with the injection of a
spin-polarized current into a superconducting wire through the F/S
interface, the superconductivity was more effectively suppressed
as the time-reversal symmetry of the superconducting pairs in the
singlet state was easily broken in the nonequilibrium state. We
estimated the spin-diffusion length $\lambda_{sp}$ from the finite
voltages revealed in the Al wire for a bias below the onset
current of superconductivity (for convenience we assign this as
the superconducting critical current), which itself was reduced by
the weakened superconductivity due to spin-polarized current
injection. The resulting spin-diffusion length saturated at
temperatures far below $T_c$ but grew gradually with increasing
temperature and tended to diverge near $T_c$. This result is
consistent with the results of Ref. 19 but is in contradiction to
the results of Refs. 20-22. The detailed temperature dependence of
$\lambda_{sp}$ in our study indicated that the spin relaxation in
a superconductor was related to the condensation of quasiparticle
pairs in two opposite spin channels into superconducting electron
pairs at the Fermi level.

\begin{figure}[b]
%h=here, t=top, b=bottom, p=separate figure page
\includegraphics[width=8 cm]{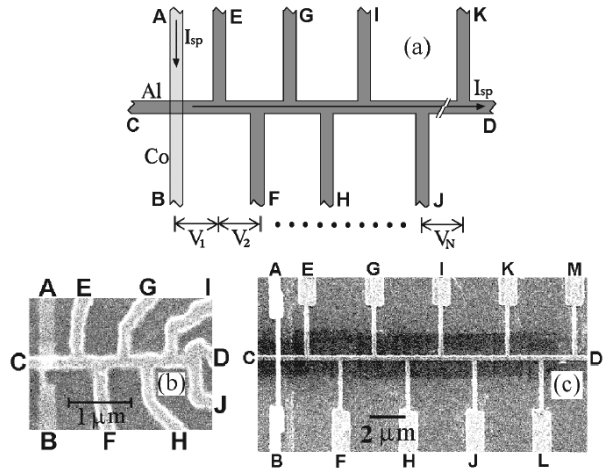}
\caption{\label{fig1}(a) Schematic geometry of the samples. SEM
micrographs of (b) the sample A and (c) the sample B.}\end{figure}

Specimens were fabricated using a combination of electron-beam
(e-beam) lithography, e-beam and/or thermal evaporation, Ar-ion
etching, and lift-off techniques. Si substrates covered with
natural oxide layers were used. For F/S hybrid samples (the
samples $A$ and $B$) ferromagnetic wires designed to form in a
single-domain structure\cite{Yunsok3} were made by the e-beam
evaporation of 60$\sim$65-nm-thick Co films on patterned layers of
e-beam resist and by lifting off subsequently to the width of
about 250$\sim$270 nm. Then about 80$\sim$130-nm-thick Al layers
for both samples with extended contact electrodes were thermally
evaporated as superconducting wires on the second patterned resist
and lifted off to the width of about 200 nm and 270 nm,
respectively. There was about 10$\%$ variation in the width of the
Al wire over the length under study for both samples. The surface
of the ferromagnetic layers was cleaned using low-energy Ar-ion
milling right before the Al deposition to enhance the transparency
of the Co/Al interface. To compare the results between the
spin-polarized and spin-degenerate configurations, a control
sample $C$ was fabricated by the same method as described above,
in which, however, the ferromagnetic Co wire was replaced by a
non-magnetic Au wire.

Schematic configuration of the samples is shown in Fig. 1(a). The
Al wire, with multiple voltage leads, was in crossed contact with
a ferromagnetic Co wire. The total number of segments of the Al
wires was 6, 9, and 6 for the samples $A$, $B$, and $C$,
respectively. For the nonequilibrium spin injection into the
superconducting Al wire the current was applied between the leads
A and D. But for the injection of spin-degenerate nonequilibrium
quasiparticles the leads C and D were used. Pair-breaking of
superconducting electrons due to the injection of the
spin-polarized current was monitored by measuring the $I-V$
characteristics of each segment of an Al wire between two
neighboring voltage leads. For the sample $A$, the voltage drop in
the segments of the Al wire $V_1$, $V_2$, $\cdot \cdot \cdot$,
$V_6$ was monitored between the leads C and E, E and F, $\cdot
\cdot \cdot \cdot \cdot \cdot$, I and J, respectively, as shown in
Fig. 1(b). The sample $C$ had the same nominal geometry as the
sample $A$. For the sample $B$ the voltage drop $V_1$, $V_2$,
$\cdot \cdot \cdot \cdot \cdot \cdot$, $V_{9}$ was also monitored
between the leads B and E, E and F, $\cdot \cdot \cdot \cdot \cdot
\cdot$, L and M, respectively, as shown in Fig. 1(c) in detail.
The center-to-center length of the segment corresponding to the
voltage drop $V_1$ (the segment one) was 460 nm (1.6 $\mu$m) and
the average center-to-center spacing between the adjacent voltage
leads for other segments was 340$\sim$380 nm (1.8 $\mu$m) for the
samples $A$ and $C$ ($B$).

Data were taken by the conventional four-probe lock-in technique
run at 38 Hz in a dilution refrigerator. The diffusion constant
$D$ of Al wire at 4.2 K, determined from the wire residual
resistivity, was 12.0 (24.8) cm$^2$/s for the sample $A$ ($B$). To
obtain the value of $D$, we used the relation\cite{Santhanam1} for
Al $\rho l_e$ = 3.2 $\times 10^{-12}$ $\Omega$cm$^2$, where $\rho$
and $l_e$ are the resistivity and the elastic mean-free path,
respectively, of the Al wires in the normal state. Here, the value
of the Fermi velocity for Al\cite{Ashcroft1} $v_F =2.03 \times
10^{8}$ cm/s was used. The interfacial resistance $R_t$ for the
sample $A$ ($B$, $C$) was about 2.4 (2.4, 0.04) $\Omega$ far below
the superconducting transition temperature $T_c$ of Al. The
corresponding interfacial transparency $t$ of the sample $A$ ($B$,
$C$), 0.22$\%$ (0.15$\%$, 11$\%$), was determined using the
relation\cite{Giroud1} of $R_t^{-1} = 2 N(E_F) v_F S e^2 t $.
Here, $N(E_F)$ and $v_F$ are the density of states at the Fermi
level and the Fermi velocity of Co (Au), respectively, for the
samples $A$ and $B$ ($C$). $S$ and $e$ are the cross-sectional
area of the interface and the electron charge, respectively.

\begin{figure}[b]
%h=here, t=top, b=bottom, p=separate figure page
\includegraphics[width=7cm]{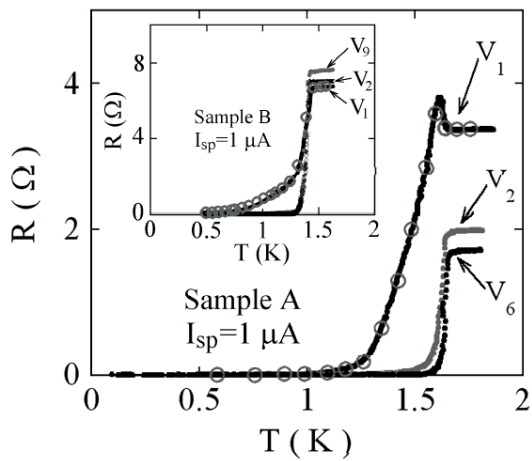}
\caption{\label{Fig2} The resistive transition of the Al wire of
the sample $A$ for different segments corresponding to the voltage
drop $V_1$, $V_2$, and $V_6$ in Fig. 1 for the bias current of 1
$\mu$A in the spin-injection configuration. Open circles are the
data corresponding to the spin-degenerate configuration. Inset:
the temperature dependence of the Al-wire resistance of the sample
$B$ for the bias current of 1 $\mu$A along the segments one, two,
and nine in the spin-injection configuration (solid curves) and
along the segment one in the spin-degenerate configuration (open
circles).}
\end{figure}

In Fig. 2 the resistance $vs$ temperature of the Al wire of the
sample $A$, determined by measuring the voltage drop $V_{1 (2,
6)}$ between the leads C (E, I) and E (F, J), is shown for a
spin-polarized bias current $I_{sp}$ of 1 $\mu$A, applied between
the leads A and D. One notes that no interfacial resistance was
included in the data in this measurement configuration. Since the
sample $A$ has a defect in the lead B near the interface [see Fig.
1(b)] this lead was not used in the measurements. The voltage drop
in the segment which is closest to the interface (the segment
one), $V_1$, shows much smeared characteristics below the onset of
the superconducting transition $T_c$ than those in other segments
(the segments two and six) such as $V_2$ or $V_6$ in the figure.
The voltage drops $V_3$, $V_4$, and $V_5$ over other segments
showed behavior (not illustrated in the figures) very similar to
$V_2$ with a few $\%$ deviation of the onset temperatures of zero
resistance. The finite resistance corresponding to $V_1$ in the
segment one below the onset of the superconducting transition is
most likely to be due to weakening of the superconductivity in the
Al wire by the spin-polarization-induced pair breaking. The
open-circle symbols are the data with the current bias of 1 $\mu$A
for the spin-degenerate bias configuration over the segment one,
where the voltage drop for the unpolarized spin injection is
almost identical to that for the case of the spin injection. This
fact indicates that the nonequilibrium effect of quasiparticle
injection is supposed to be minimal for this low bias level.

On the other hand, the identical results between the two bias
configurations imply that, even for this quasi-equilibrium
situation in the low spin-degenerate bias current, pair breaking
comparable to the level for corresponding spin injection takes
place. Random interdiffusion of conduction electrons even without
an external bias current can take place crossing the interface.
This, in turn, induces spin accumulation in the Al wire near the
interface, because the spin population of the two opposite
polarities is imbalanced in the ferromagnetic Co wire. The
resulting spin acculmulation in the superconducting Al wire
induces the pair breaking and causes the finite resistance below
the bulk transition temperature $T_c$ of Al. Thus, the finite
resistance below $T_c$ of the Al wire is not due to the
bias-induced pair breaking but is due to the self spin injection
near the interface. This is similar to the ``self injection"
effect as discussed in Ref. 27. The difference in the normal-state
resistance for different segments resulted from the variation in
the length as well as in the width of segments. The unusual peak
in the resistance corresponding to $V_1$ is presumably due to
nonuniform current distribution at the junction as the Al
electrode became superconducting. This peak feature appeared even
in the Au/Al junction of the sample C.

As illustrated in the inset of Fig. 2 similar behavior was
observed in the wire resistance $vs$ temperature of the sample $B$
for the segments represented by $V_1$, $V_2$, and $V_9$. For the
sample $B$ also the open-circle data corresponding to $V_1$ for
the spin-degenerate bias configuration are almost the same as
those for the spin injection configuration. This indicates again
that the bias level of 1 $\mu$A used to determine the temperature
dependence of resistance of the sample $B$ was low enough so that
the equilibrium electron state in the Al wire was not disturbed
even for the spin-injection bias configuration. The spatial
dependence of the resistance in Fig. 2 also reveals that the
spin-polarized state of the bias current was confined within the
segment one of the Al wire in both samples.

\begin{figure}[b]
%h=here, t=top, b=bottom, p=separate figure page
\includegraphics[width=6.8 cm]{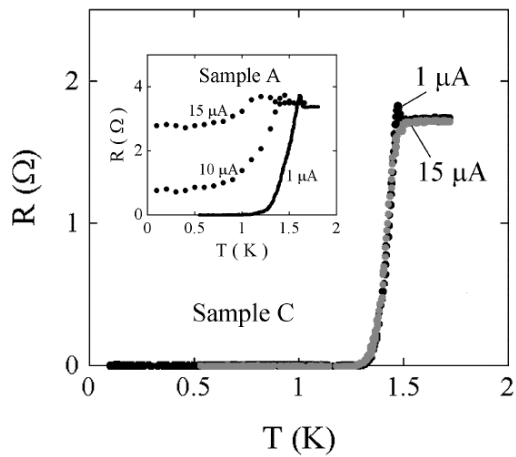}
\caption{\label{Fig3} The resistive transition for the segment one
of the Al wire of the sample $C$, consisting of a Au/Al junction,
for the bias currents of 1 and 15 $\mu$A. Inset: the resistive
transition of the segment one of the Al wire of the sample $A$ for
increasing spin-polarized bias current from 1 to 15 $\mu$A.}
\end{figure}

\begin{figure}[b]
%h=here, t=top, b=bottom, p=separate figure page
\includegraphics[width=7cm]{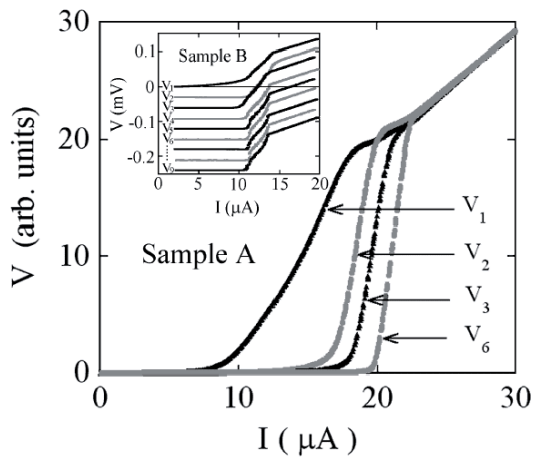}
\caption{\label{Fig4} $I-V$ characteristics of the sample $A$
taken at 0.10 K, along the segments one, two, three, and six of
the Al wire for the spin-injection bias configuration. Inset: the
spatial dependence of the $I-V$ characteristics taken from the
segments 1, 2, ....., 9 of the Al wire of the sample $B$ at 0.43
K. For clarity each curve is offset downward from the nearest
neighbor by 0.03 mV.}
\end{figure}

\begin{figure}[b]
%h=here, t=top, b=bottom, p=separate figure page
\includegraphics[width=7.5cm]{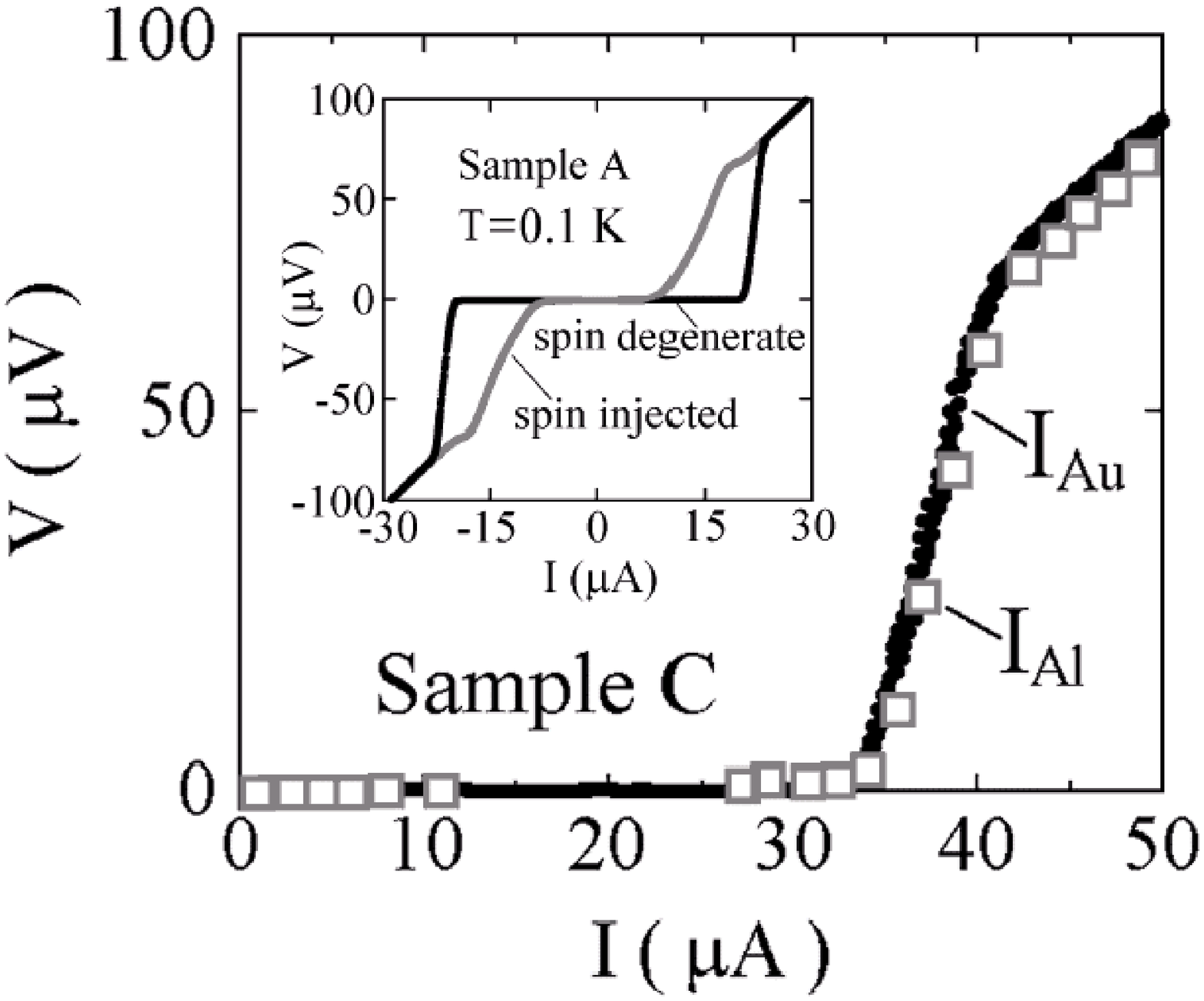}
\caption{\label{Fig5} $I-V$ characteristics for the segment one in
the control sample $C$ (consisting of Au/Al junction), taken at
0.10 K for the bias current fed from Au lead (solid circle) and
from Al lead (open square), which would correspond to
spin-polarized and spin-degenerate configurations, respectively,
in the samples $A$ and $B$. In the inset the $I-V$ characteristics
for the segment one in the sample $A$ at 0.10 K in the
spin-injection configuration (the grey curve) in comparison with
the spin-degenerate bias configuration (the black curve).}
\end{figure}

The behavior of the Al-wire resistance that was almost insensitive
to the bias between spin-injection and spin-degenerate
configurations changed for higher current biases. The inset of
Fig. 3 shows again the resistance $vs$ temperature of the segment
one of the Al wire of the sample $A$ for increasing spin-polarized
bias current from 1 to 15 $\mu$A. For the bias of 10 $\mu$A a
finite resistance appeared even below the original value of $T_c$,
which indicates that, for this bias level, significant
spin-polarization-induced pair breaking took place. For 15 $\mu$A
almost full pair breaking is visible. In comparison, for the
spin-degenerate bias configuration, the resistive transition of
the Al wire for the samples A and B remained almost unaltered for
the current bias up to 15 $\mu$A (the data are not shown). On the
other hand, when we injected a current through a nonmagnetic Au
wire, no noticeable pair breaking effect was visible up to 15
$\mu$A for any bias modes. Fig. 3 shows such resistive transition
for the segment one of the Al wire of the sample C. In this sample
consisting of Au/Al junction the transition of the segment one of
the Al wire is much sharper than in the previous case consisting
of Co/Al interface. Apparently in this case no pair breaking due
to spin accumulation effect dominated the resistive-transition
characteristics of the Al wires.

Fig. 4 shows the spatial dependence of the $I-V$ characteristics
of the segments one, two, three, and six of the sample $A$
measured at 0.10 K in the spin-polarized bias configuration. The
voltage value of each segment is normalized with respect to the
normal-state resistance. Except for small variation the segments
two, three, and six show transition to the normal state at
corresponding critical currents with almost equal sharpness. In
contrast, the transition of the segment one is much smeared with a
significantly reduced critical current. The appearance of the
clear finite resistance in the segment one below its critical
current is due to the pair breaking by the spin injection. As
observed in the resistive-transition data in Fig. 2, the spatial
variation of the $I-V$ curve also indicates that the spin
injection effect decays within the range comparable to the length
of the segment one of superconducting Al wire.

In the inset of Fig. 4 we also illustrate the spatial dependence
of the spin injection effect exhibited in the $I-V$
characteristics of the sample $B$. Different sets of $I-V$
characteristics were taken from the segments one, two, $\cdot$
$\cdot$ $\cdot$, and nine\cite {sample B} at 0.43 K. For clarity,
each set is offset downward from the neighboring curve by 0.03 mV.
In this sample also the finite voltage below the critical current
is present only for the segment one, which is consistent with the
picture that it was caused by the pair breaking due to the
nonequilibrium spin injection within the spin-diffusion length
near the interface.

The inset of Fig. 5 clearly contrasts with the $I-V$
characteristics of the segment one of the sample $A$ measured at
0.1 K between the two different configurations: the grey curve
shows the characteristics for the spin-injection configuration and
the black curve is the one without spin injection. For the
spin-injection configuration the $I-V$ curve is much smeared with
a significantly reduced critical current. The slightly peaked
feature in the voltage near the critical current above the
normal-state value in the spin-injection configuration is not well
understood. But the feature appeared only in the segment one so
that one may assume it was caused by nonuniform current
distribution at the junction.

We took the nonequilibrium conduction properties of the Al wire in
a sample where the ferromagnetic Co wire was replaced by
non-magnetic normal wire, $i.e.$, the sample $C$. In this case the
injected current was spin degenerate in any bias configurations.
In the main panel of Fig. 5, $I-V$ characteristics of the segment
one of the control sample $C$ are compared between biasing through
leads A and D as denoted by $I_{Au}$ and biasing through leads C
and D as denoted by $I_{Al}$, which would correspond to the
spin-polarized and spin-degenerate mode, respectively, for the
samples A and B. $I-V$ characteristics turn out to be almost
identical in both bias configurations, because pair breaking due
to spin injection was absent in both cases. Slight discrepancy
between the two curves arose from the possible difference in the
effective length of the segment one between the two configurations
and/or the nonuniform current distribution at the interface for
the bias current of $I_{Au}$. Even for this spin-degenerate
configuration, however, pair breaking by the nonequilibrium
current injection may have smeared the superconducting transition
of the Al wire near the critical current as seen in the figure.

One may argue that the seeming spin-injection effect was caused by
simple Joule heating generated by a bias current in the
ferromagnetic wire or at the interface. In fact, the control
sample $C$ where the seeming spin-injection effect was absent had
a interfacial resistance much lower than the samples $A$ and $B$
with Co/Al interfaces. In order to interpret the suppression of
superconductivity described above in terms of spin-related pair
breaking one need to rule out the possibility of thermally induced
pair breaking effect. To examine the possibility of Joule heating
at the interface conduction properties of Al wire in a sample with
much higher interfacial resistance were measured. Fig. 6 shows the
differential resistance measured in another test sample at
temperatures far below $T_c$,
%\cite{Yunsok2}
over three different distances from the interface. The junction
area of this sample was similar to that of the rest of the samples
and the interfacial resistance of this sample was 17.4 $\Omega$,
almost an order of magnitude higher than the samples A and B. One
notices that all the curves, including the one for the segment one
that is closest from the interface, have similar sharpness of the
transition with almost the same values of the critical current. If
there were significant contribution of heating at the interface
the segment one should show much smeared characteristics with a
reduced critical current. The behavior of the curves in this
figure indicates that the heating effect is supposed to be
insignificant even for a junction with resistance much higher than
those of the samples A and B. On the other hand, in this test
sample with higher interfacial resistance, the spin injection is
supposed to be ineffective because of the spin flip scattering at
the interface. Thus, the spin injection effect was not present in
the data of Fig. 6. This argument indicates that the appearance of
finite voltages below the critical currents in the spin-injection
configuration, in the samples A and B, resulted from pair breaking
due to spin injection to the Al wires both with the relatively low
interfacial resistance of 2.4 $\Omega$.

\begin{figure}[b]
%h=here, t=top, b=bottom, p=separate figure page
\includegraphics[width=8.5cm]{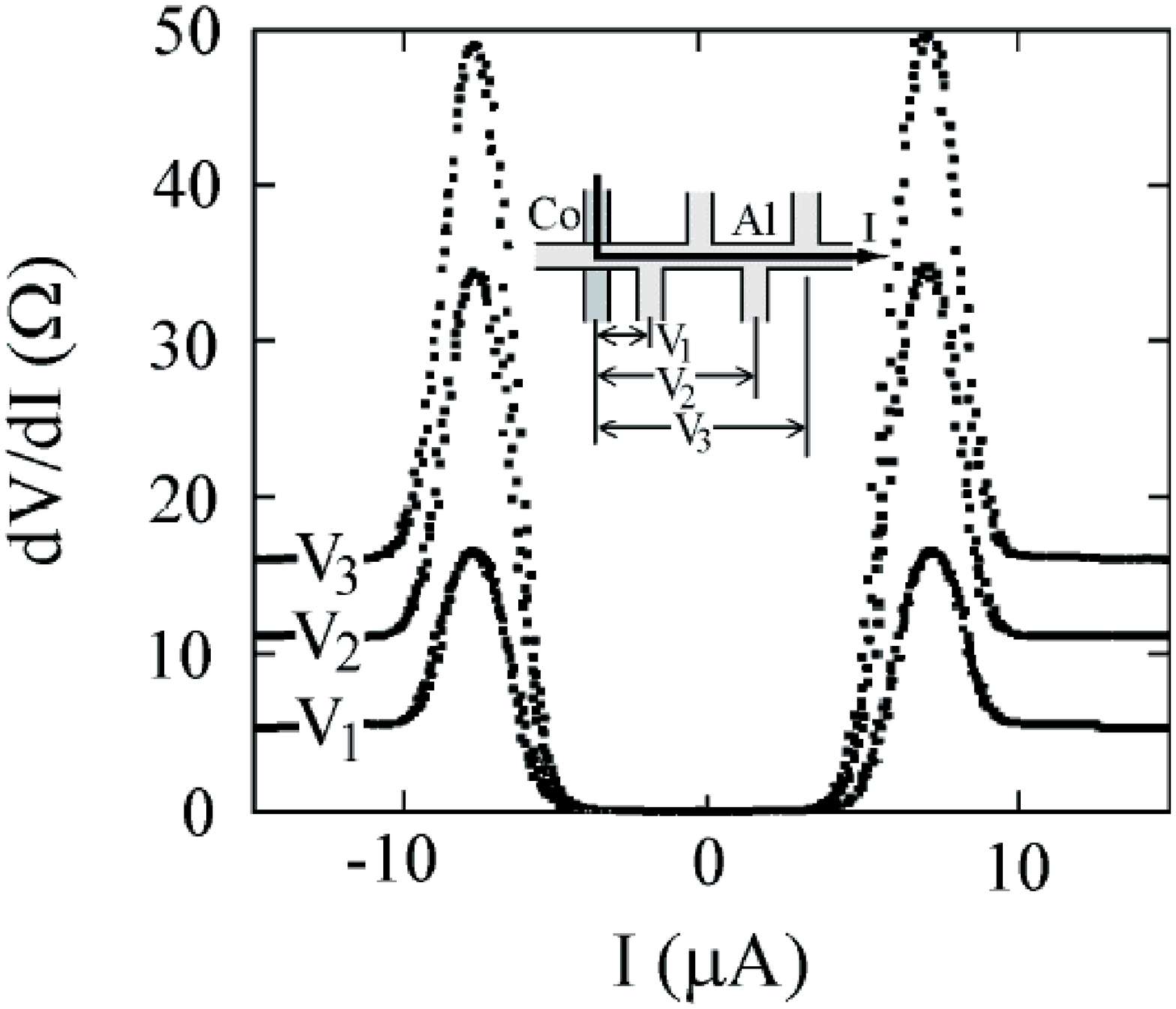}
\caption{\label{Fig6} The differential resistance measured in a
test sample consisting of Co/Al interface at temperatures far
below $T_c$ over three different distances from the interface.}
\end{figure}

\begin{figure}[b]
%h=here, t=top, b=bottom, p=separate figure page
\includegraphics[width=7cm]{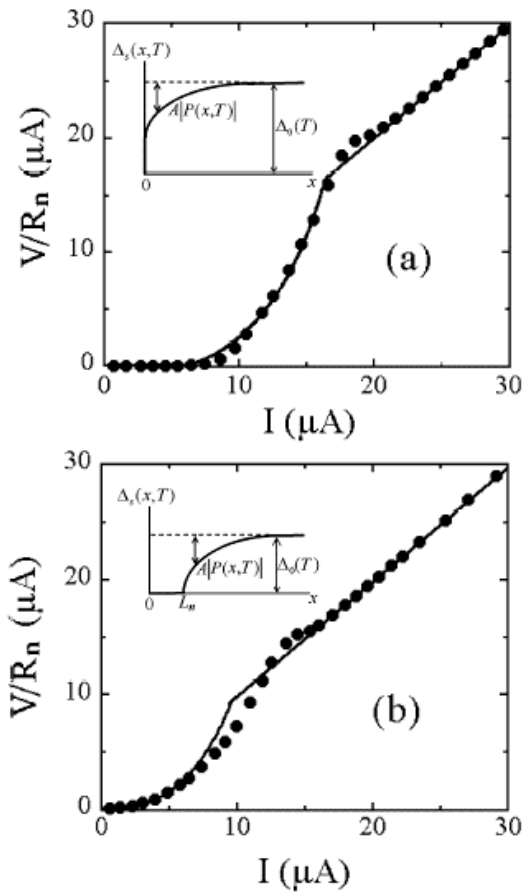}
\caption{\label{Fig7} $I-V$ characteristics (open circles) of the
segment one for temperatures (a) far below $T_c$ ($T$ = 0.1 K) (b)
and near $T_c$ ($T$ = 1.3 K) in the sample $A$, with the best-fit
curves (solid curve) using Eqs. (2) and (3).}
\end{figure}

\begin{figure}[b]
%h=here, t=top, b=bottom, p=separate figure page
\includegraphics[width=8cm]{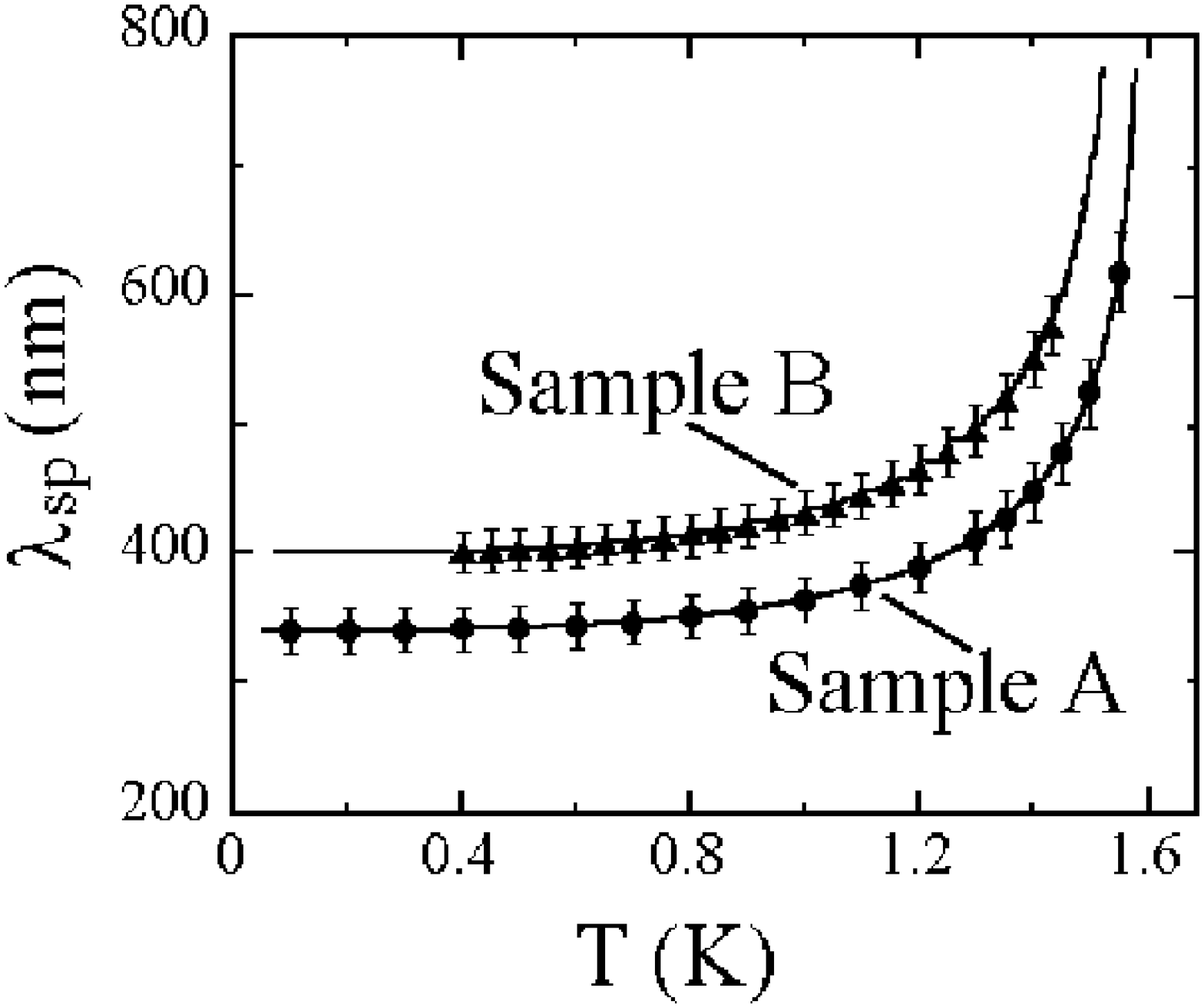}
\caption{\label{Fig8} The temperature dependence of $\lambda_{sp}$
for the samples $A$ (circles) and $B$ (triangles), extracted from
the best-fit curves in $I-V$ characteristics based on Eqs. (2) and
(3). The solid curves are the best fits to the relation
$\lambda_{sp}=\sqrt{D \tau_{sp}}$, together with Eq. (4).}
\end{figure}

We estimate the effective spin diffusion length $\lambda_{sp}$
from the finite voltages below the critical current by adopting a
phenomenological model. Suppose a superconducting wire is placed
along the $x$ axis with the F/S interface at $x$=0. In the model,
local superconducting gap $\Delta_s (x,T)$, in the presence of the
spin accumulation near the interface of F/S, is assumed to be
$\Delta_0 (T) - A |P (x,T)|$ for $\Delta_0 (T)
> A |P (x,T)|$ and zero otherwise. Here, $\Delta_0 (T)$ is the
local superconducting gap in the absence of the spin accumulation,
$|P (x,T)|$ is the absolute density of the spin imbalance, and $A$
is a parameter defined as (a dimensionless
constant)$\times$$1/N_n$, where $N_n$ is the density of states per
unit volume in the normal state. The local critical current $I_c
(x,T)$ is assumed to be $B \Delta_s (x,T)$, where $B$ is another
parameter defined as (a dimensionless constant)$\times$$N_n e
v_F$$\times$(the cross section of a superconducting wire). Then,
the voltage drop $V$ over a region of Al wire of length $L$ from
the interface for an applied current $I$ is given by
\begin{eqnarray}
V & = & \int_0^L dx \frac{\Delta V}{\Delta x}\nonumber\\
  & = & \int_0^L dx I R_n \frac{1}{L} \theta(I-I_c(x,T))\nonumber\\
  & = & I R_n \frac{1}{L} \int_0^L dx \theta(I-I_c(x,T))\nonumber\\
  & = & I R_n \frac{L_n^{eff}}{L},\label{Eq1}
\end{eqnarray}
where $\theta(y)$ is the step function, which is 1 for $y > 0$ and
0 otherwise. Here, $R_n$ and $L_n^{eff}$ are the resistance of the
Al wire and the effective spin diffusion length in the normal
state, respectively. The total voltage drop $V$ is the sum of the
local voltage drop $\Delta V$ over an infinitesimal segment
$\Delta x$. The local voltage drop $\Delta V$ appears when the
applied bias current $I$ exceeds the local critical current $I_c
(x,T)$ of an infinitesimal segment $\Delta x$ located at $x$. From
the assumption above, the critical current $I_c (L_n^{eff},T)$ is
determined by the relation $I_c =B [\Delta_0 (T) - A |P
(L_n^{eff},T)|]$. If the local density of spin accumulation is
assumed to relax exponentially as $P(x,T)=P_0(T)
$exp$[-x/\lambda_{sp}(T)]$ the effective spin diffusion length
follows the relation, $L_n^{eff}=\lambda_{sp} $log$[A B P_0 / (B
\Delta_0 - I)]$. Hence, the voltage drop $V$ is obtained as
\begin{eqnarray}
V & = & 0, \ \ \ {\rm for} \ 0 < I  < B \Delta_0 - A B P_0 \nonumber\\
  & = & I R_N, \ \ \ {\rm for} \ I > B \Delta_0 \nonumber\\
  & = & I R_N \frac{\lambda_{sp}}{L} log [\frac{A B P_0}{B \Delta_0 -
I}], \ \ \ {\rm otherwise}. \label{Eq2}
\end{eqnarray}

This relation is satisfied for a strong superconducting state with
large $\Delta_0 (T)$ in the temperature range sufficiently below
$T_c$. In this case the spatial distribution of the
superconducting strength may look like the one as illustrated in
the inset of Fig. 7(a). As the temperature approaches $T_c$,
however, a certain range over the length $L_n$ of the Al wire from
the interface loses the superconductivity with vanishing
$\Delta_s(x,T)$ as $\Delta_0$ becomes smaller than $A |P(x,T)|$
near $T_c$ [see the inset of Fig. 7(b)]. Then, the spatial
dependence of $P(x,T)$ for $x>L_n$ is modified as $P_0(T)
$exp$[-L_n/\lambda_n(T)] $exp$[-(x-L_n)/\lambda_{sp}(T)]$. Here,
$L_n$ and $\lambda_n$ are the length of normal region for
$\Delta_0 < A |P(x,T)|$ and the spin diffusion length in the
normal state, respectively. The ratio of $L_n / L$ is assumed to
be proportional to the ratio between the zero-bias-limit
resistance and the normal-state resistance near $T_c$. In this
case, the voltage drop $V$ is also modified as
\begin{eqnarray}
V & = & I R_n \{ \frac{L_n}{L} + \frac{\lambda_{sp}}{L}
log[\frac{A B P'_0}{B \Delta_0 - I}] \}, \  {\rm for} \ 0 < I < B \Delta_0 \nonumber\\
  & = & I R_n, \  {\rm otherwise} .
\label{Eq3}
\end{eqnarray}
where, $P'_0$ = $P_0(T) $exp$[-L_n/\lambda_n(T)]$.

Using Eqs. (\ref{Eq2}) and (\ref{Eq3}), the spin diffusion lengths
far below $T_c$ and near $T_c$ are extracted, respectively. We
adopted three fitting parameters $\lambda_{sp}$, $ A B P_0$ and $B
\Delta_0$ for the best fit to Eq. (2). $ A B P_0$ should be less
than $B \Delta_0$ and the value $B \Delta_0 - A B P_0$ is the
maximum bias current of the zero-resistance state in the
temperature regime far below $T_c$. On the other hand, we adopted
two parameters $\lambda_{sp}$ and $ A B P'_0$ for the best fit to
Eq. (3). The value of $A B P_0$ must be larger than $B \Delta_0$
in the temperature range near $T_c$. In the fit the value of $A B
P_0$ near $T_c$ is extracted from the value of the quantity for
$T\ll T_c$ as obtained in the fit to Eq. (2), while assuming a
linear temperature dependence. $B \Delta_0$ near $T_c$ is also
determined from its value far below $T_c$ incorporated with the
BCS-type temperature dependence of the energy gap, $\Delta_0
(T)$.\cite{Xu}

As discussed in relation with Eq. (2), $I-V$ curves at 0.10 K in
the sample $A$ show the three different characteristic regimes of
voltage drop $V$ for a range of bias current $I$: the zero
resistance regime, the finite-voltage regime below the critical
current and the normal resistance regime above the critical
current. In the finite-voltage regime, the three fitting
parameters, $\lambda_{sp} = 340$ nm, $ A B P_0 = 14$ $\mu$A and $B
\Delta_0 = 20$ $\mu$A at 0.10 K, are determined from the best fit
(solid line) to the $I-V$ curves in Fig. 7(a). It turns out,
however, that the quality of the best-fit curve is not much
sensitive to the fitting parameter values within 10 $\%$ of
variation. The resulting best-fit parameter values give the
relative magnitudes among parameters that are consistent with the
assumptions given above. In comparison, in Fig. 7(b), the $I-V$
curves at 1.3 K show two regimes of voltage drop $V$: the
finite-voltage regime below the critical current and the
normal-resistance regime above the critical current. The features
in Figs. 7(a) and 7(b) are consistent with the assumed variation
of the superconducting strength as illustrated in their insets in
relation with Eqs. (2) and (3), respectively. The length of
normal-state region $L_n$ at 1.3 K, as estimated from the
zero-bias-limit resistance, is 48 nm. The best-fit values (solid
line) of the parameters turn out to be $\lambda_{sp} = 410$ nm and
$ A B P'_0 = 11$ $\mu$A. In this fit we used the local gap value,
corresponding to $ B \Delta_0 = 13.7$ $\mu$A , obtained from the
BCS behavior.

The value $ A B P_0 = 11.2$ $\mu$A at 1.3 K, which is obtained by
linearly extrapolating the low-temperature-limit values as
obtained from the fit in relation with Fig. 7(a), is not in
agreement with the assumption of $ A B P_0> B \Delta_0$. This
contradiction presumably originates from the naive assumptions of
step function in Eq. (1) and/or the linear dependence between the
critical current and the energy gap. One may believe that the
existence of the zero-bias-limit resistance implies $\Delta_s = 0$
at the interface, but the fitting formula of Eq. (3) may hold only
approximately in the intermediate temperature range between 0 and
$T_c$. The fit, following the same procedure, to $I-V$
characteristics far below $T_c$ and near $T_c$ for the sample $B$
gave similar quality of the fit (not shown).

In Fig. 8 we plot the temperature dependence of $\lambda_{sp}$
extracted from the best-fit to $I-V$ characteristics. It shows
that the spin diffusion length $\lambda_{sp}$ is almost
temperature independent in the temperature range far below $T_c$,
which is 1.6 K (1.56 K) for the sample $A$ ($B$). The
zero-temperature-limit value of $\lambda_{sp}(0)$ for the sample
$A$ $(B)$ was 340 nm (400 nm). The empirical value of
$\lambda_{sp}$ increases with $T$ and tends to diverge near $T_c$.
This temperature dependence of $\lambda_{sp}$ turns out to be in
remarkable agreement with that observed in the $c$-axis
spin-polarized quasiparticle tunneling in
YBa$_2$Ca$_3$O$_{7-\delta}$ thin films.\cite{Yeh} The temperature
dependence of $\lambda_{sp}$ is also in qualitative agreement with
the results obtained in Nb\cite{Johnson1} but in clear
contradiction with result in Refs. 20 and 21, where $\lambda_{sp}$
decreases for temperatures approaching $T_c$. The temperature
dependence of $\lambda_{sp}$ also contradicts to the theoretical
results of Ref. 22, where the spin-diffusion length is predicted
to be the same both in the normal and in the superconducting
states, implying that the spin diffusion length in a
superconductor should be almost independent of temperature in the
range of our study.

The spin diffusion length in the normal state in our study is
estimated to be $\lambda_n \sim$ 1 $\mu$m from the ratio between
the extrapolated value of $P_0(T)$ and the fitting parameter of
$P'_0(T)$, with 50 $\%$ variation in its value in the temperature
range near $T_c$ where the assumption of $A B P_0 > B \Delta_0$ is
satisfied. Thus, the temperature dependence of $\lambda_n$ cannot
be accurately determined near $T_c$. The spin relaxation time in
the normal-metallic state $\tau_n$ in the sample $A (B)$ is
calculated to be about 450 (1170) ps at 1.4 K using the relation
of $\lambda_n$ = $\sqrt{D \tau_n}$, which is in comparison with
the previous results\cite{Jedema3} for $\tau_n$ of 100 ps at 4.2 K
obtained using the nonlocal spin-injection measurements.

%%%%%%%%%%%%%%%%%%%%%%%%%%%%%%%%%%%%%%%%%%%%%%%%%%%%%%%%%%%%%%%%%%
% ( $\lambda_n$ in the sample $A (B)$ is also estimated as about 350
% (500) nm from the previous results\cite{Jedema3} for the spin
% relaxation time $\tau_n$ of 100 ps at 4.2 K using the relation of
% $\lambda_n$ = $\sqrt{D \tau_n}$, which is in agreement with the
% results in our study. )
%%%%%%%%%%%%%%%%%%%%%%%%%%%%%%%%%%%%%%%%%%%%%%%%%%%%%%%%%%%%%%%%%%%

Employing the picture of the relaxation of charge-imbalanced
nonequilibrium quasiparticle states in a
superconductor,\cite{Schmid} the spin relaxation time has been
suggested to follow the relation,\cite{Yeh}
\begin{equation}
\tau_{sp} \sim \tau_{ex} k_B T_c /\Delta (T).
\end{equation}
Here, the energy-relaxation time or the inelastic-scattering time
$\tau_{ex}$ is defined in terms of the spin exchange as $\tau_{ex}
\sim \hbar /h_{ex}$ ($h_{ex}$ is the exchange energy inside the
superconductor) and $\Delta (T)$ is the superconducting energy
gap. In this picture, the nonequilibrium spin imbalance is set by
the characteristic energy-relaxation or inelastic scattering time
but only the fraction of quasiparticles ~$\delta/k_B T_c$ just
above the gap is effectively involved in relaxing the spin
imbalance.\cite{Schmid} Then temperature dependence of the spin
diffusion length, expressed as $\lambda_{sp}=\sqrt{D \tau_{sp}}$,
should be determined by the temperature dependence of $\Delta$ as
$1/\sqrt{\Delta (T)}$. The best fit to this temperature dependence
is shown for the samples $A$ and $B$ in Fig. 8 by solid curves. In
the fit we use the empirical formula\cite {Xu} $\Delta (T)
=\Delta$(0)tanh$(1.74 \sqrt{T_c/T -1})$ for the temperature
dependence of the gap, which is supposed to be valid in all the
temperature range below $T_c$ [=1.6 (1.56) K], with $T_c$ as the
fitting parameter for the sample $A$ $(B)$. Combining
$\lambda_{sp}(0)$= 340 (400) nm with $D$=12.0 (24.8) cm$^2$/sec
for the sample $A$ $(B)$, the spin relaxation time in the Al wire
for $T \ll T_c$ is estimated to be $\tau_{sp} \sim 9.6$
$(6.5)\times 10^{-11}$ sec for the sample $A$ $(B)$. The
corresponding exchange energy $h_{ex} /k_B$ for the sample $A$
$(B)$ was 91 mK (95 mK), which is larger than the value of 11 mK
for Nb.\cite{Johnson1} The fast spin relaxation, corresponding to
the large exchange energy, in Al was discussed in Ref. 9, in terms
of the pseudopotential band calculation results by Fabian and Das
Sarma.\cite {Fabian} It is theoretically suggested that the small
spin hot spots at the large Fermi surface of polyvalent metals
like Al give excessive contribution to the spin flip scattering,
making the spin relaxation faster by up to a factor of 100. The
nice fit of the temperature dependence of $\lambda_{sp}$, on the
other hand, indicates that the spin diffusion in superconductors
is governed by the energy relaxation between the opposite spin
channels as well as the pair condensation over the superconducting
gap.

The spin-relaxation length measured previously in the normal state
of Al\cite{Jedema4} at 4.2 K was 1200 nm, which is thus longer
than that in the superconducting state by a factor of $\sim$ 4 as
measured in this study. Although the direct comparison of the spin
diffusion lengths in systems with different electron diffusivity
is meaningless the above trend may indicate that the spin
diffusion length in the normal state is, in general, longer than
that in the superconducting state. One may explain this trend in
terms of plausible spin-relaxation processes in superconducting
system in the following way. An imbalanced nonequilibrium state of
the spin-polarized quasiparticles between the opposite spin bands
in the superconductor, caused by the spin injection, relaxes to a
non-equilibrium spin-balanced state, which in turn relaxes to the
equilibrium condensed Cooper-paired state. The (second)
recombination process in a superconductor depopulates the
quasiparticles in the nonequilibrium state, which expedites the
(first) spin-flip process mediated by the spin-orbit interaction.
We believe that is why the spin-relaxation in the superconducting
state is more effective than that in the normal state. We thus
suppose the fast increase of the spin diffusion length near $T_c$
should be limited by its normal-state value, although it could not
be confirmed in our study because of the lack of the resolution in
the measurements of the spin diffusion length very close to $T_c$.

It is surprising that a large spin-injection effect was observed
in spite of the rather small interfacial transparency in the
samples A and B. As pointed out in Ref. 37, the spin injection
rate through the interface of low transparency is proportional to
the interfacial polarization and the ratio between the interfacial
resistance and the resistance corresponding to the spin-diffusion
length in the non-magnetic electrode. The interfacial polarization
decreases with increasing interfacial resistance in a system with
a diffusive interface as the interfacial spin-flip scattering
occurs more frequently. But, at the same time, the ratio between
the two resistance values increases with increasing the
interfacial resistance. We suppose the two competing factors kept
the spin-injection efficiency high enough in our systems of finite
interfacial resistance close to 2.4 $\Omega$. A quantitative
estimate of the spin-injection rate, however, is not available
because the first-principle calculation of the interfacial
polarization with spin-flip scattering is not available.

In conclusion, we observed the suppression of the nonequilibrium
superconductivity, induced by spin-polarized quasiparticle
injection into mesoscopic superconducting Al wires in proximity
contact with an overlaid ferromagnetic Co wire. The suppression,
as evidenced by the occurrence of finite voltages for the
bias-current range below the superconducting onset, was pronounced
when the spin-polarized currents were injected through the Co/Al
interfaces. The finite voltages in the samples with transparent
interfaces of low interfacial resistances are attributed to the
dynamic pair breaking by the quasiparticles with the imbalanced
spin population. The temperature dependence of the spin diffusion
length in a superconductor, estimated from the finite voltages
over a certain length of Al wire near the interface, suggests that
the spin diffusion in the superconductor is governed by the pair
condensation of quasiparticles through opposite spin channels.
Since the pair condensation depopulates the spin-balanced
quasiparticles more efficient spin flip can take place, via the
spin-orbit interaction, in the superconducting state than in the
normal state, making the spin diffusion length, in general,
shorter in the superconducting state.

\section*{Acknowledgments}
This work was supported by Electron Spin Science Center, in Pohang
University of Science and Technology, administered by KOSEF. This
work was also partially supported by Nano Research and Development
Program administered by KISTEP.


\begin{thebibliography}{00}
\bibitem{sample B} Once in this study the spin injection effect
was observed over several $\mu$m range in the superconducting Al
wire. Multiple leads in the sample B were arranged over an
extended length of the Al wire to examine the seeming long-range
character of the spin diffusion. But it has not been reproduced
ever since so that the extended multiple-leads arrangement in the
sample B turned out to be unnecessary. We believe the strange
behavior was caused by some defects in the sample.
\end{thebibliography}
\end{document}